\documentclass[conference]{IEEEtran}

\usepackage{amsmath,amssymb,amsfonts}
\usepackage{booktabs}
\usepackage{adjustbox}
\usepackage{algorithm}
\usepackage{longtable}
\usepackage{array}
\usepackage{multirow}
\usepackage[caption=false,font=normalsize,labelfont=sf,textfont=sf]{subfig}
\usepackage{textcomp}
\usepackage{bibentry}
\usepackage{bibentry}
\usepackage{stfloats}
\usepackage{url}
\usepackage{verbatim}
\usepackage{graphicx}
\usepackage{cite}
\usepackage[pdftex]{pdfpages}
\usepackage{xcolor}
\usepackage[nice]{nicefrac}
\usepackage[colorlinks,bookmarksopen,bookmarksnumbered,citecolor=blue,urlcolor=blue]{hyperref}
\usepackage{algpseudocode}

\begin{document}
\newcommand{\BIBentryALTinterwordspacing}{%
  \spaceskip=\fontdimen2\font plus \fontdimen3\font minus \fontdimen4\font\relax}
\newcommand{\BIBentrySTDinterwordspacing}{\spaceskip=0pt\relax}
\title{No Attention, No Problem: DPU-Aware Attention Approximation in Modern YOLO on FPGA}



\author{Suraj Karki, Qazi Arbab Ahmed*\footnote{*}, Thorsten Jungeblut \\
\IEEEauthorblockA{Bielefeld University of Applied Sciences and Arts (HSBI), Bielefeld, Germany
}
Email: \{suraj.karki, qazi.ahmed, thorsten.jungeblut\}@hsbi.de
}

\renewcommand{\thefootnote}{\fnsymbol{footnote}}
\markboth{Journal of \LaTeX\ Class Files,~Vol.~14, No.~8, August~2021}%
{Shell \MakeLowercase{\textit{et al.}}: A Sample Article Using IEEEtran.cls for IEEE Journals}


\maketitle
\thefootnote{}
\footnotetext{%
*Corresponding author
}
\begin{abstract}
Edge-based Artificial Intelligence (AI) acceleration has recently improved progress in real-time object detection. Object detection on edge devices requires a balance between accuracy, speed and power efficiency. This paper proposes a customized Deep Learning Processor Unit (DPU)-aware architecture for attention-based YOLO variants deployed on AMD FPGAs. Specifically, we evaluate and benchmark YOLOv26 and YOLOv11—two modern attention-based YOLO variants—on the Xilinx ZCU104 across both standard and oriented object detection tasks.


We replace unsupported activation functions, substitute split operations with 1x1 convolutions, and approximate the spatial attention mechanism in a DPU-compatible way. All models are then trained and evaluated across six benchmark datasets---COCO, Pascal VOC, KITTI, DOTA, DIOR-R, and an in-house human presence dataset---and benchmarked across all eight DPU configurations (B512 to B4096) in terms of mAP, FPS, latency, power, and resource utilization. Notably, YOLOv26n and YOLOv26n-OBB deliver the highest end-to-end throughput at 34.05 and 29.55 FPS for standard and oriented detection, respectively, with an average of $5$\%  absolute reduction in accuracy due to quantization, while achieving up to approximately $3{\times}$ lower power consumption compared with the state-of-the-art.

\end{abstract}

\begin{IEEEkeywords}
 FPGA, Object Detection, YOLO, Oriented Bounding Box, Vitis AI, Edge Computing, Deep Learning Accelerator, Quantization.
\end{IEEEkeywords}

\section{Introduction}
\label{sec:Intro}
Object detection is the foundation of modern computer vision, which enables autonomous systems to identify and locate objects within images or video streams \cite{MANI2022105319}\cite{DiCecio_2024}. It plays an important role across various domains, including medical imaging, autonomous driving, and surveillance systems. As these applications often require fast and reliable results, improving algorithm efficiency, speed, and accuracy remains a key research focus \cite{inproceedings}. Many specialized domains—especially remote sensing—require rotated bounding boxes for object detection to accurately capture targets at different rotations and scales, which regular horizontal boxes can't do well \cite{Wang_2025}. Meeting these accuracy and efficiency requirements for both standard and oriented detection relies on the detector architecture.

To create a robust object detector, different techniques have been developed which enhance detection algorithms based on deep learning. Convolutional Neural Networks (CNNs) have given object detection algorithms the power to perform better \cite{ad}. There are generally two types of modern object detectors: two-stage detectors and single-stage detectors. The two-stage detector first generates candidate regions and then refines bounding boxes while classifying objects in a second stage. These methods achieve very high precision but are computationally expensive and have high latency. This results in less suitability for real-time applications \cite{inproceedings}. In comparison to a one-stage detector, it treats the problem as a single-shot regression problem and makes predictions in a single step. It makes it fast, but it might have slightly lower accuracy compared to two-stage detectors \cite{make5040083}. YOLO (You Only Look Once) models have emerged as the state-of-the-art for this category, which has several versions. The architectures have evolved to balance accuracy with speed and ensure efficient execution on hardware-constrained devices \cite{ad}.


Field-Programmable Gate Arrays (FPGAs) offer a significant advantage over traditional CPUs and GPUs due to their ability to provide high performance with low power consumption \cite{10039025}.  They are experiencing rapid growth in the domain of AI acceleration, driven by their capacity for parallel processing and architectural optimizations. Unlike fixed-logic processors, FPGAs offer hardware-level reconfigurability, which allows developers to modify circuit logic and wiring specifically for neural network parallelisms or even switch models dynamically at runtime \cite{Nguyen2024}. This architecture enables faster execution through high parallelism and custom data-paths, which is critical for real-time applications such as autonomous driving, drone navigation, and human presence detection \cite{electronics14203993}. To achieve these benefits, modern detectors and their components, such as attention mechanisms, must be adapted for commercial FPGA toolchains.

Even though plenty of studies have been done on FPGA-based accelerators, there is a lack of end-to-end deployment of state-of-the-art versions of the YOLO family across both standard (horizontal) and oriented object detection tasks on an FPGA System-on-Chip (SoC). This paper presents a systematic study of adapting state-of-the-art YOLO-based detectors for AMD Vitis AI deployment, including architecture adaptations for attention-based variants, quantization, compilation, and onboard benchmarking. The main contributions of the paper are summarized as follows:
\begin{enumerate}
  \item We present a customized DPU-aware attention approximation architecture for modern object detection algorithms such as YOLOv26 and YOLOv11 for efficient deployment on an FPGA SoC using commercial tools.
  \item We benchmark the end-to-end performance of these models on an FPGA across standard object detection and oriented object detection using state-of-the-art datasets.
  \item We evaluate performance across different architectures of the DPUCZDX8G IP core, providing a detailed analysis of latency, FPS, power, and hardware resource consumption. 
  \item Our work provides a quantitative baseline for deploying standard and oriented bounding box object detection models for various applications on an FPGA SoC using the Vitis AI toolchain.
\end{enumerate}

The remaining parts of this paper are organized as follows. Section \ref{sec:relatedwork} summarizes previous literature work on object detection algorithms and their implementation in FPGAs. Section \ref{sec:PropMeth} describes the proposed methodology used for this study. Section \ref{sec:results} presents the study's results. Section \ref{sec:Con} summarizes the work and indicates future directions for the research.

\section{Related Work}
\label{sec:relatedwork}
Object detection is a fundamental task in computer vision, which has made significant advancements over the years. This section reviews the different YOLO architectures and existing work on FPGA-based acceleration of YOLO models.

\subsection{Object Detection}
Early two-stage detectors, such as Faster R-CNN, achieved good detection quality, but carried a computational overhead that makes real-time edge deployment difficult \cite{ren2016fasterrcnnrealtimeobject, e24101346}. YOLO, introduced in $2016$, is a pioneer of single-stage detectors. The model predicts bounding boxes and class probabilities simultaneously on a single pass by treating them as a direct regression problem \cite{redmon2016lookonceunifiedrealtime}. This opened the door to real-time inference on resource-constrained devices. The YOLO family has developed considerably over time: early versions used Darknet backbones with anchor-based prediction, and later, they introduced anchor-free detection from YOLOv8, efficient feature fusion, and attention mechanisms. YOLOv11 added the C3k2 block and a spatial attention module (C2PSA), and YOLOv26 eliminated distribution focal loss and non-maximum suppression (NMS) to further reduce complexity and post-processing overhead \cite{li2020generalizedfocallosslearning, chakrabarty2026yolo26analysisnmsfreeend}.

\subsection{FPGA Acceleration Of YOLO Models}
FPGAs have demonstrated a suitable inference platform for edge-based AI applications. Graphical Processing Units (GPUs) offer high computational power but consume significant power, whereas FPGAs offer reconfigurable architectures that enable parallel processing with lower power consumption \cite{10.1145/3020078.3021740}, which makes them well-suited for YOLO-based object detection.
 
Early studies focused on lightweight YOLO variants on low-end FPGAs. Ngo et al. \cite{10378810} deployed a $12$-layer YOLOv2 on a Cyclone V SoC using OpenCL-based General Matrix Multiply (GEMM) acceleration and achieved $90$\% accuracy at $4$ FPS with only $32$K LUTS. Bao et al. \cite{bao} replaced the standard convolution operation with the Winograd convolution algorithm in YOLOv2 to deploy on the PYNQ-Z2 board, reducing DSP usage and achieving $124$ ms latency at $2.7$W. Lee et al. \cite{10930191} extended this by resolving Winograd's stride-$2$ reordering inefficiency via a row-stacking technique, then applying adaptive  INT8 asymmetric quantization to reach $194.7$ FPS at $1.57$W with under $1$\% accuracy loss on the ZCU106 SoC. Ding et al. \cite{ding2019reqyoloresourceawareefficientquantization} developed REQ-YOLO for Tiny-YOLO, which uses the Alternating Direction Method of Multipliers (ADMM) to automatically determine optimal bit-widths across layers, achieving $314.2$ FPS on a Virtex-7 with only $0.07$\% mAP loss – showing that quantization strategy matters as much as a target bit-width. Valadanzoj et al. \cite{Valadanzoj2024} developed a high-speed YOLO accelerator specific for autonomous driving applications on an FPGA by applying a genetic algorithm to optimize fixed-point decimal positions across layers and achieved $55$ FPS at $79$\% accuracy and $13.6$W.


Vitis AI is a widely adopted commercial toolchain for deploying deep learning models on Xilinx SoC platforms. Wang et al. \cite{wang} deployed YOLOv3 on the ZCU104, achieving $206.7$ FPS at $3.38$ FPS/W, while Amin et al. \cite{amin} reported comparable results with YOLOv3-Tiny on the Kria KV260 for traffic light detection, reaching $99$\% accuracy at $3.5$ FPS/W. Nguyen et al. \cite{Nguyen2024} achieved $29$ FPS on the ZCU104 for aerial object detection with only $1$\% accuracy drop, and Liu et al. \cite{Liu_2024} deployed YOLOv5 for flame detection at $56$ FPS and $4.147$W with a $1.64$\% accuracy loss.

A streaming data-flow architecture, such as FINN~\cite{finn}, is also ideal for its ultra-low latency performance, where each layer is mapped to a dedicated hardware module with continuous data transfer between stages. Calì et al. \cite{electronics14203993} applied the FINN framework to generate a dataflow accelerator for YOLOv3-Tiny, achieving $6$ ms latency at $2.55$W, but with degradation in mAP from $41.76$\% to $17.7$\% due to aggressive quantization. Machura et al. \cite{jlpea12020030} compared the performance of Vitis AI, FINN, and custom-built accelerators on the Ultra96-V2 and found that a custom design achieved the highest throughput of $196$ FPS, while FINN achieved $111$ FPS through a dataflow architecture, and the Vitis AI DPU offered the most automated workflow, confirming a throughput-versus-convenience trade-off. Montgomerie-Corcoran et al. \cite{montgomeriecorcoran2023sataystreamingarchitecturetoolflow} developed the SATAY toolflow, which automatically generates a streaming architecture for YOLO models (YOLOv3 to YOLOv8) across multiple FPGA SoC boards. Danilowicz and Kryjak~\cite{danilowicz2025realtimemultiobjecttrackingusing} deployed a nano YOLOv8 for multi-object tracking on the ZCU102 using the FINN framework, reaching $195.3$ FPS at the FPGA level but only approximately $24$ FPS end-to-end due to software overhead. 

Prior work on FPGA-based object detection has focused predominantly on earlier YOLO variants, leaving modern attention-based architectures such as YOLOv11 and YOLOv26 underexplored. 
Moreover, oriented object detection on FPGA remains largely unexplored. This work addresses both gaps through a systematic benchmark of YOLOv8, YOLOv11, and YOLOv26 – including OBB variants – across state-of-the-art datasets and all eight DPUCZDX8G architectures on the Xilinx ZCU104.

\section{Methodology}
\label{sec:PropMeth}
\begin{figure}[h]
    \centering
    \includegraphics[width=\linewidth]{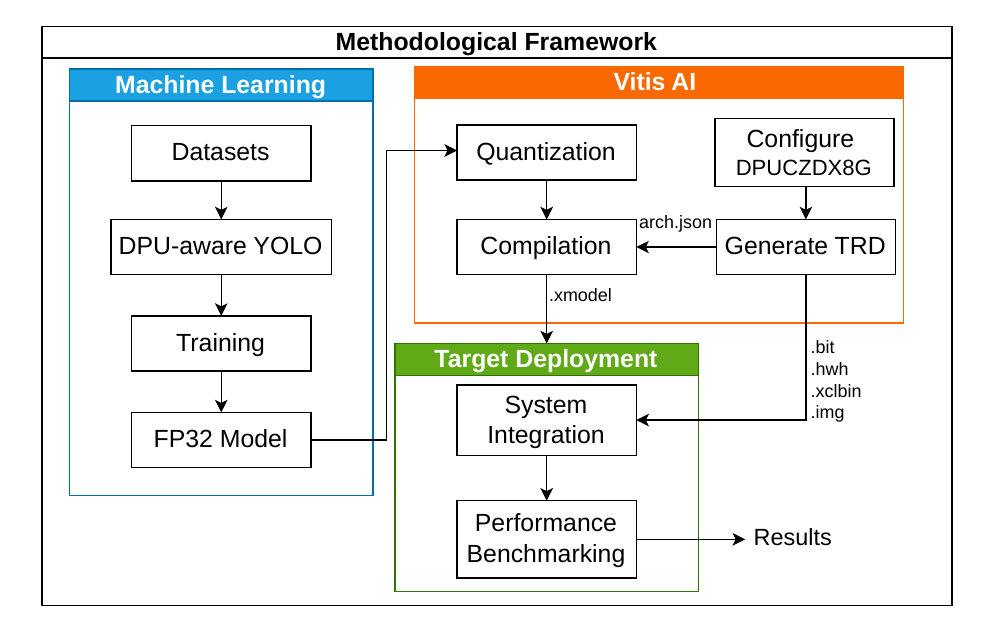}
    \caption{The proposed methodological framework.}
    \label{fig:methodological_framework}
\end{figure}

This section consists of dataset preparation, model customization, training and quantization, compilation, and deployment on the target platform. Fig. \ref{fig:methodological_framework} shows the methodological framework we used for this study.
\vspace{-3mm}
\subsection{Dataset}
The performance of customized YOLO architectures is benchmarked using state-of-the-art datasets. COCO, Pascal VOC, and KITTI are used for standard object detection tasks, whereas DOTA and DIOR-R are used for oriented object detection. A custom-built person detection dataset is used for both tasks. The datasets are formatted according to the Ultralytics framework.

The Pascal VOC dataset contains $21,503$ images across $20$ object classes and serves as a standard benchmark for object detection and visual object category recognition \cite{Everingham2010}. The COCO dataset \cite{lin2015microsoft} consists of $200$k images with $80$ object categories and approximately $1.5$ million annotated instances, making it one of the most widely used benchmarks for object detection evaluation. The KITTI dataset \cite{Geiger2013IJRR} has $7,481$ stereo camera images annotated across $8$ object categories, and is commonly used as a standard benchmark for autonomous driving and computer vision applications. The DIOR-R dataset \cite{Cheng_2017}, \cite{mandal2025diorr} is a transformation of the DIOR dataset for oriented object detection, which contains $23,463$ high-resolution $1024$ × $1024$ aerial images annotated with $20$ object categories. DOTA \cite{9560031} is another benchmarking dataset for oriented object detection, which contains $13,955$ images, preprocessed for YOLO at a fixed $1024$ × $1024$ via resizing, splitting, and padding where necessary. A custom person detection dataset is built from CCTV footage and VOC dataset images (single class: person). It contains $18,820$ images at $416$ × $416$ and $184,789$ annotations. The dataset is split into $15,056$ training, $1,882$ validation, and $1,882$ test images. Data augmentation is applied, including horizontal flip ($50$\%), random rotation (±$15$°), shear (±$7$°), and brightness/exposure adjustments (±$10$\%).

\subsection{DPU-Aware Approximations of YOLO}
This study implemented the nano variants of YOLOv8, YOLOv11, and YOLOv26, including both standard and oriented object detection models. This section describes the necessary DPU-aware adaptations made to the models to be deployed on the AMD FPGA using Vitis AI, as certain operators, activation functions, and classes are not supported by the DPU. Without these changes, unsupported components execute on the CPU, resulting in a reduction of the overall performance. The Ultralytics \cite{ultralytics_yolo} library is adapted to integrate the required modifications to the model architecture and training pipeline.

\subsubsection{YOLOv8 and YOLOv8-OBB}
\label{subsubsec:y8}
\begin{figure}[h]
    \vspace{-3mm}
    \centering
    \includegraphics[width=0.85\linewidth]{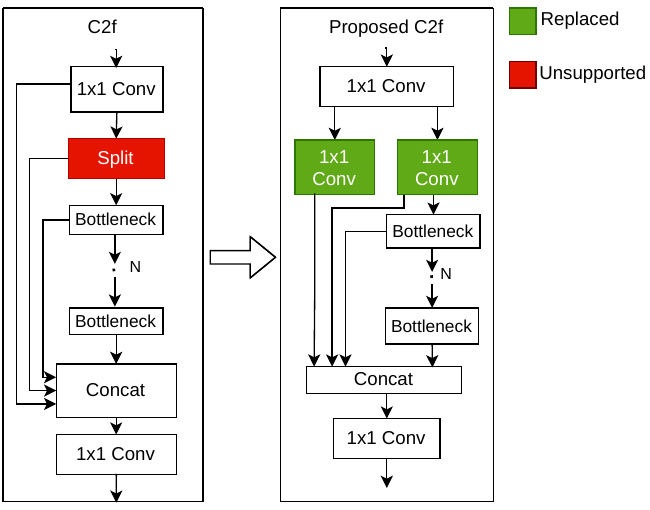}
    \caption{Original C2f block (left) and the proposed block with two $1 \times 1$ convolution blocks (right).}
    \label{fig:c2f_replacement}
\end{figure}

YOLOv8 is an anchor-free, one-stage detector that predicts a probability distribution over the four bounding box parameters per pixel, with the final box obtained by averaging those predictions \cite{li2020generalizedfocallosslearning}. Since the Xilinx DPU does not support the Sigmoid Linear Unit (SiLU) activation function, which is present on both YOLOv8 and YOLOv8-OBB, we replaced it with the Rectified Linear Unit (ReLU) activation function. The split operation in the feature-fusion block (C2f) is unsupported, so it is replaced with equivalent linear projection operations. This is done through a $1 \times 1$ convolution operation, which is supported by a DPU. A $1 \times 1$ convolution acts as a per-pixel linear projection across channels. We initialize the $1 \times 1$ convolution weights as a binary selection matrix: diagonal entries are set to $1$ for the required channel mapping, and all other entries are set to 0. Each output channel copies exactly one input channel, effectively reproducing the split. In this way, two $1 \times 1$ convolutions can extract the same two-channel groups that a split would produce, which can be seen in Fig. \ref{fig:c2f_replacement}.
\subsubsection{YOLOv11 and YOLOv11-OBB}
\label{sec:yolov11}
YOLOv11 retains the overall structure of YOLOv8 but adds key upgrades. It replaces the previous C2f block with the C3k2 block, a more efficient CSP bottleneck that improves detection accuracy and accelerates feature extraction. The model also performs multi-scale detection using three heads that process neck feature maps to produce final classification and localization outputs.\cite{khanam2024yolov11overviewkeyarchitectural}.

\begin{figure}[h]
    \centering
    \includegraphics[width=\linewidth]{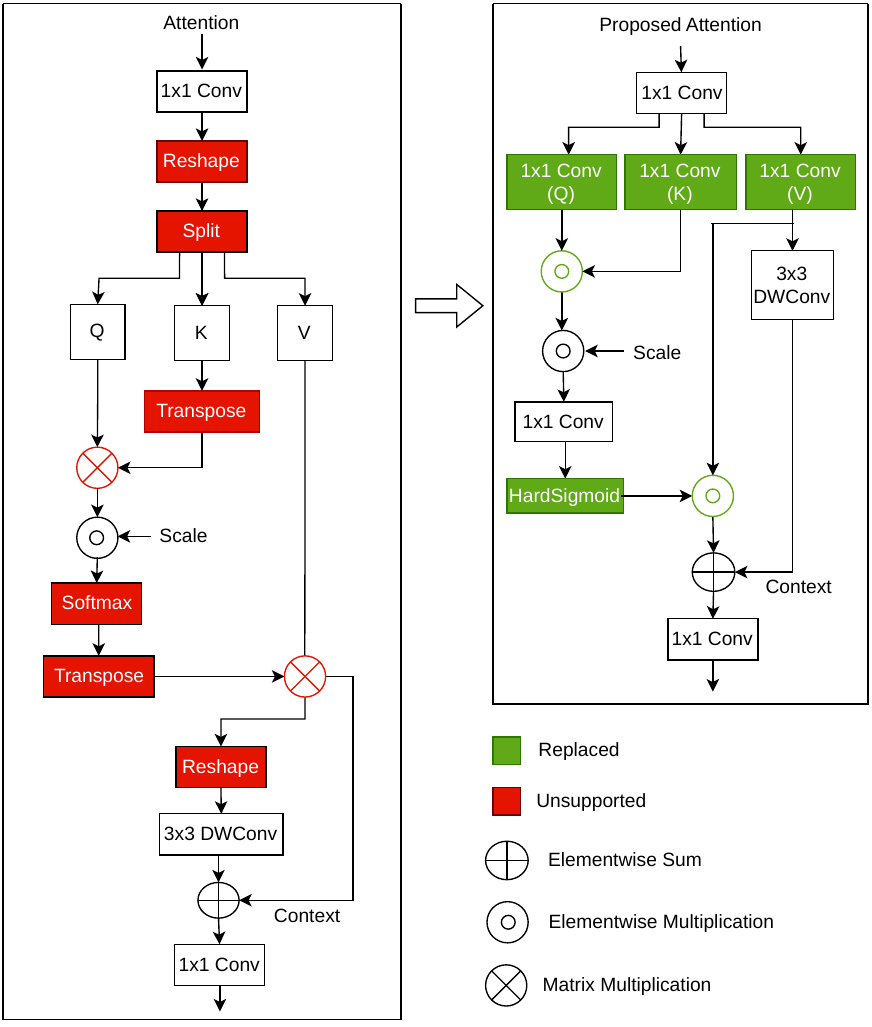}
    \caption{The original attention block (left) and the proposed DPU-aware attention block (right).}
    \label{fig:attention_replacement}
\end{figure}

The most important concept introduced in YOLOv11 is spatial attention through the Cross-Stage Partial Spatial Attention (C2PSA) module. The attention mechanism enables the model to focus on important regions within the feature map, which results in more robust object detection. However, the entire attention block cannot be executed on the FPGA because reshape, transpose, matrix multiplication, and softmax operations are not natively supported by the Xilinx DPU. To overcome these hardware constraints, we propose a DPU-compatible attention approximation implemented using standard convolution and element-wise operations, shown in Fig. \ref{fig:attention_replacement}.
In this implementation:

\begin{enumerate}
    \item The Query ($q$), Key ($k$), and Value ($v$) features are extracted using parallel $1 \times 1$ convolutions.
    \item The attention weights are approximated via a scaled element-wise product ($q \odot k$).
    \item The softmax activation function is replaced with the hard-sigmoid function to support normalization.
    \item The result is projected back to the value channels through a $1 \times 1$ convolution before gated fusion with the residual context.
\end{enumerate}

\begin{figure}[h]
    \centering
    \includegraphics[width=0.90\linewidth]{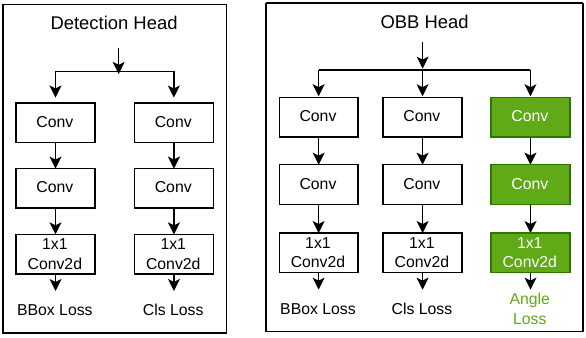}
   \caption{Standard Detection Head and Oriented Bounding Box (OBB) Head.}
    \label{fig:head}
\end{figure}

We utilize the YOLOv11-OBB variant for oriented object detection. The architecture remains mostly similar to the standard version, but the detection head includes an additional regression branch to predict the rotation angle ($\theta$) of bounding boxes, shown in Fig. \ref{fig:head}. This enables the model to precisely localize tilted objects, which is critical for certain applications such as aerial imagery, industrial inspection, surveillance, etc.

\subsubsection{YOLOv26 and YOLOv26-OBB}
The YOLOv26 architecture is designed with the aim of an efficient edge deployment philosophy. Most of its features are retained from YOLOv11, but it introduces two major improvements: the removal of distribution focal loss (DFL) and a head supporting end-to-end NMS-free inference. The DFL is replaced with a direct bounding box regression, which predicts continuous values directly (x, y, w, h, $\theta$) \cite{sapkota2026yolo26keyarchitecturalenhancements}. This study still uses NMS to benchmark accuracy across all the datasets. To accelerate the model on FPGA, the architecture is modified using the same ReLU replacement, $1 \times 1$ convolution mapping, and DPU-compatible attention approximation detailed in Section \ref{sec:yolov11}.

\subsection{Model Training and Compression}
The Ultralytics library is employed for training YOLO models because of its simplicity, speed, and strong out-of-the-box performance. All YOLO variants are trained with similar hyperparameter settings to ensure consistency and fairness across model comparisons. Ultralytics automatically selected the appropriate optimizer based on the number of training iterations. An initial learning rate of $0.01$ was applied, decayed to a final factor of $0.01$ following a cosine annealing schedule. A momentum of $0.937$ and a weight decay of $0.0005$ were applied to regularize training and stabilize convergence. A linear warm-up phase spanning $3$ epochs was incorporated at the beginning of training to gradually ramp up the learning rate prior to the main optimization process. The input image was scaled to $416$ pixels for the longer edge. The models were initialized using pre-trained checkpoints to leverage transfer learning and accelerate convergence. Mixed-precision training (FP16) was enabled throughout all experiments to reduce memory consumption and improve computational efficiency without compromising model accuracy.

There are two types of quantization: post-training quantization (PTQ) and quantization-aware training (QAT).
In this work, we used the PTQ method, as it is sufficient to achieve 8-bit quantization with accuracy similar to that of $32$-bit. The process involves three steps: ($1$) pre-processing the model by fusing batch normalization layers with convolutions and removing irrelevant nodes; ($2$) calibrating the model on a subset of data to estimate quantization parameters (scale and zero-point); and ($3$) converting the model to an $8$-bit integer representation in a DPU-compatible format for deployment.

\subsection{Model Compilation}
The quantized model is compiled for the ZCU104 SoC board using the Vitis AI compiler, targeting the DPUCZDX8G IP. Initially, the compiler transforms the quantized model into the Xilinx Intermediate Representation (XIR), a graph-based format, acting as a key intermediate format to connect AI models with hardware by enabling quantizers, compilers, and runtimes to optimize and execute them efficiently. The graph is then divided into smaller subgraphs, where optimization is applied. The compiler also generates instruction streams for the DPU parts of the model. Finally, all necessary information is packaged for the Vitis AI runtime (VART), and the model is compiled into the .xmodel format for deployment. During the compilation process, the proper fingerprint of the DPUCZDX8G IP is specified to make the compiled model compatible with it. Vitis AI offers various architectures (B512, B800, B1024, B1152, B1600, B2304, B3136, B4096) for the DPUCZDX8G IP core based on the level of hardware parallelism and available FPGA resources. This enables a trade-off analysis between performance and hardware utilization for the model to be deployed.

\subsection{Hardware Platform and VITIS TRD Flow}\label{sec:vitistrd}
The Vitis Targeted Reference Design (TRD) and the pre-built, unmodified DPUCZDX8G IP core provided by AMD for the ZCU104 SoC are used to rebuild the hardware design using Vivado (v2022.2) and Vitis (v2022.2). The design is rebuilt for eight different DPUCZDX8G IP core architectures (B512 through B4096) by editing the existing configuration parameters in $\texttt{dpu\_conf.vh}$, enabling channel augmentation, URAM, high RAM usage, and low-power mode, and re-running the standard TRD build flow. After building the project for each architecture, the $\texttt{arch.json}$, bitstream ($\texttt{.bit}$), hardware design ($\texttt{.hwh}$), device binary ($\texttt{.xclbin}$), and deployable binary image ($\texttt{.img}$) files are generated. These files are used to boot the target board with the built DPU architecture from the SD card.

\subsection{Program Flow}
\begin{figure}[h]
    \centering
    \includegraphics[width=0.90\linewidth]{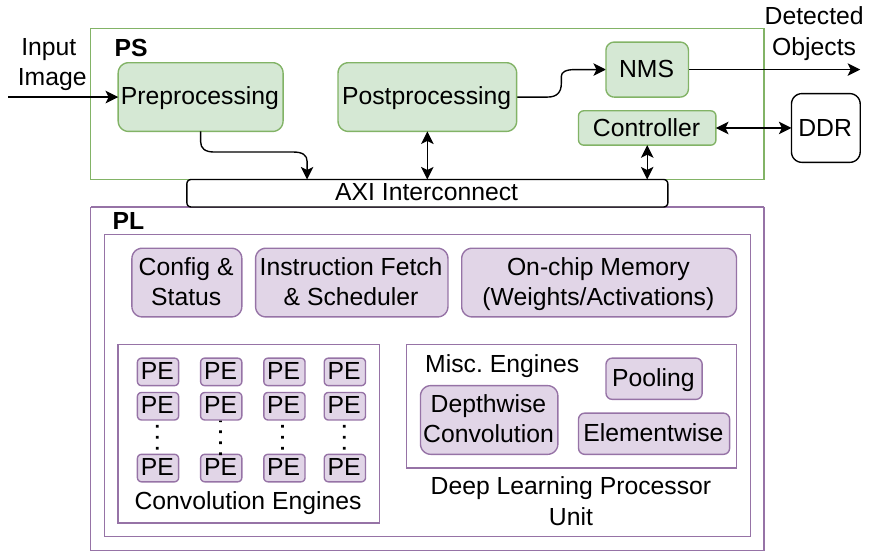}
   \caption{Implementation of the hardware-software system on the ZCU104 SoC FPGA target.}
    \label{fig:ps-pl}
\end{figure}

This is the final step in deploying the compiled model. To run the model on an FPGA SoC board, a software application is required to manage communication between the Processing System (PS) and the Programmable Logic (PL), which contains the DPU. The application is responsible for reading input images from the PS, performing pre-processing, passing the pre-processed data to the DPU on the FPGA, and then fetching the DPU output to perform post-processing as shown in Fig. \ref{fig:ps-pl}. Pre-processing consists of scaling the input image to the appropriate size, normalizing the pixels, and sending it to the accelerator in the PL. During postprocessing, the three feature maps obtained from the three detection heads are transformed into bounding boxes, and redundant bounding boxes are removed using the NMS process. The application is developed in Python using the Vitis AI Library, which provides APIs for communication between the DPU and the ARM processor.
\vspace{-5mm}

\section{Results}
\label{sec:results}
\subsection{Experimental Setup}
An Ubuntu $24.04.4$ LTS host machine with Intel(R) Core i7-13800H and $32$GB of RAM is used for model preparation, quantization, and compilation, while model training is performed on high-performance GPU clusters. The Xilinx ZCU104 Evaluation Kit is used as the target deployment and benchmarking platform. The platform is equipped with an AMD Zynq UltraScale+ MPSoC EV device, which contains a quad-core ARM Cortex-A53 applications processor, $16$nm FinFET+ programmable logic, making it well-suited for embedded computer vision applications.

\subsection{Performance Evaluation}
We evaluated the attention-approximated models after each step for end-to-end benchmarking using the mean average precision (mAP) metric, specifically mAP@$0.5$:$0.95$, which averages precision across ten IoU thresholds to comprehensively measure detection and bounding box localization accuracy. 
As shown in Table~\ref{tab:combined_original_modified}, the proposed modifications to the original models result in an average reduction of only $3$\% (absolute) in mAP@$0.5$:$0.95$ across most datasets for all detectors compared to baseline models. 
After deployment on the FPGA, the models experience an average reduction of approximately $7$\% in mAP@$0.5$:$0.95$ across datasets for attention-based YOLO variants as shown in Table~\ref{tab:combined_cpu_fpga}. This degradation is mainly due to the quantization of the models from $32$-bit floating-point to $8$-bit integer representation.
Overall, YOLOv11 and YOLOv11-OBB frequently achieve the highest performance across all datasets.
\begin{table*}
\centering
\caption{mAP@$0.5$:$0.95$ of Original and Modified YOLO Models for Standard and Oriented Object Detection using different datasets.}
\label{tab:combined_original_modified}
\begin{tabular}{
l
c c@{\hspace{8pt}}
c c@{\hspace{8pt}}
c c@{\hspace{8pt}}
c c
}
\toprule
\multicolumn{9}{l}{\textbf{Standard Object Detection}} \\
\midrule
\multirow{2}{*}{\textbf{Model}} &
\multicolumn{2}{c}{\textbf{VOC}} &
\multicolumn{2}{c}{\textbf{KITTI}} &
\multicolumn{2}{c}{\textbf{COCO}} &
\multicolumn{2}{c}{\textbf{Person\_Detection}} \\
\cmidrule(lr){2-3}\cmidrule(lr){4-5}\cmidrule(lr){6-7}\cmidrule(lr){8-9}
& \textbf{Original} & \textbf{Modified}
& \textbf{Original} & \textbf{Modified}
& \textbf{Original} & \textbf{Modified}
& \textbf{Original} & \textbf{Modified} \\
\midrule
YOLOv8n
& $0.60$ & $0.60$
& $0.63$ & $0.63$
& $0.33$ & $0.27$
& $0.75$ & $0.73$ \\
YOLOv11n
& $0.61$ & $0.59$
& $0.63$ & $0.60$
& $0.35$ & $0.31$
& $0.77$ & $0.75$ \\
YOLOv26n
& $0.61$ & $0.57$
& $0.62$ & $0.58$
& $0.36$ & $0.31$
& $0.75$ & $0.71$ \\
\midrule
\multicolumn{9}{l}{\textbf{Oriented Object Detection}} \\
\midrule
\multirow{2}{*}{\textbf{Model}} &
\multicolumn{2}{c}{\textbf{DIOR-R}} &
\multicolumn{2}{c}{\textbf{DOTA}} &
\multicolumn{2}{c}{\textbf{Person\_Detection}} &
\multicolumn{2}{c}{} \\
\cmidrule(lr){2-3}\cmidrule(lr){4-5}\cmidrule(lr){6-7}
& \textbf{Original} & \textbf{Modified}
& \textbf{Original} & \textbf{Modified}
& \textbf{Original} & \textbf{Modified}
& & \\
\midrule
YOLOv8n-OBB
& $0.66$ & $0.66$
& $0.43$  & $0.43$
& $0.84$ & $0.84$
& & \\
YOLOv11n-OBB
& $0.62$ & $0.61$
& $0.44$ & $0.34$
& $0.86$ & $0.83$
& & \\
YOLOv26n-OBB
& $0.62$ & $0.60$
& $0.42$ & $0.32$
& $0.85$ & $0.81$
& & \\
\bottomrule
\end{tabular}
\end{table*}

\begin{table*}[t]
\centering
\caption{Performance comparison of YOLO models across multiple datasets on CPU and FPGA, measured in terms of mAP@$0.5$:$0.95$.}
\label{tab:combined_cpu_fpga}
\setlength{\tabcolsep}{5pt}
\renewcommand{\arraystretch}{1.12}
\begin{adjustbox}{max width=\textwidth}
\begin{tabular}{
l
c c@{\hspace{10pt}}
c c@{\hspace{10pt}}
c c@{\hspace{10pt}}
c c
}
\toprule

\multicolumn{9}{l}{\textbf{Standard Object Detection}} \\
\midrule
\multirow{2}{*}{\textbf{Model}} &
\multicolumn{2}{c}{\textbf{VOC}} &
\multicolumn{2}{c}{\textbf{KITTI}} &
\multicolumn{2}{c}{\textbf{COCO}} &
\multicolumn{2}{c}{\textbf{Person\_Detection}} \\
\cmidrule(lr){2-3}\cmidrule(lr){4-5}\cmidrule(lr){6-7}\cmidrule(lr){8-9}
& mAP$_{0.5:0.95}^{\mathrm{CPU}}$ & mAP$_{0.5:0.95}^{\mathrm{FPGA}}$
& mAP$_{0.5:0.95}^{\mathrm{CPU}}$ & mAP$_{0.5:0.95}^{\mathrm{FPGA}}$
& mAP$_{0.5:0.95}^{\mathrm{CPU}}$ & mAP$_{0.5:0.95}^{\mathrm{FPGA}}$
& mAP$_{0.5:0.95}^{\mathrm{CPU}}$ & mAP$_{0.5:0.95}^{\mathrm{FPGA}}$ \\
\midrule
YOLOv8n
& $0.60$ & $0.45$
& $0.64$ & $0.34$
& $0.27$ & $0.26$
& $0.73$ & $0.66$ \\
YOLOv11n
& $0.59$ & $0.57$
& $0.60$ & $0.59$
& $0.31$ & $0.29$
& $0.75$ & $0.71$ \\
YOLOv26n
& $0.57$ & $0.48$
& $0.58$ & $0.51$
& $0.31$ & $0.29$
& $0.71$ & $0.67$ \\

\midrule

\multicolumn{9}{l}{\textbf{Oriented Object Detection}} \\
\midrule
\multirow{2}{*}{\textbf{Model}} &
\multicolumn{2}{c}{\textbf{DIOR-R}} &
\multicolumn{2}{c}{\textbf{DOTA}} &
\multicolumn{2}{c}{\textbf{Person\_Detection}} &
\multicolumn{2}{c}{} \\
\cmidrule(lr){2-3}\cmidrule(lr){4-5}\cmidrule(lr){6-7}
& mAP$_{0.5:0.95}^{\mathrm{CPU}}$ & mAP$_{0.5:0.95}^{\mathrm{FPGA}}$
& mAP$_{0.5:0.95}^{\mathrm{CPU}}$ & mAP$_{0.5:0.95}^{\mathrm{FPGA}}$
& mAP$_{0.5:0.95}^{\mathrm{CPU}}$ & mAP$_{0.5:0.95}^{\mathrm{FPGA}}$
& & \\
\midrule
YOLOv8n-OBB
& $0.66$ & $0.56$
& $0.43$ & $0.31$
& $0.84$ & $0.75$
& & \\
YOLOv11n-OBB
& $0.61$ & $0.53$
& $0.34$ & $0.30$
& $0.83$ & $0.79$
& & \\
YOLOv26n-OBB
& $0.60$ & $0.52$
& $0.32$ & $0.29$
& $0.81$ & $0.77$
& & \\
\bottomrule
\end{tabular}
\end{adjustbox}
\end{table*}
\subsection{FPS and Latency Analysis}
\begin{figure*}
    \centering
    \includegraphics[width=\linewidth]{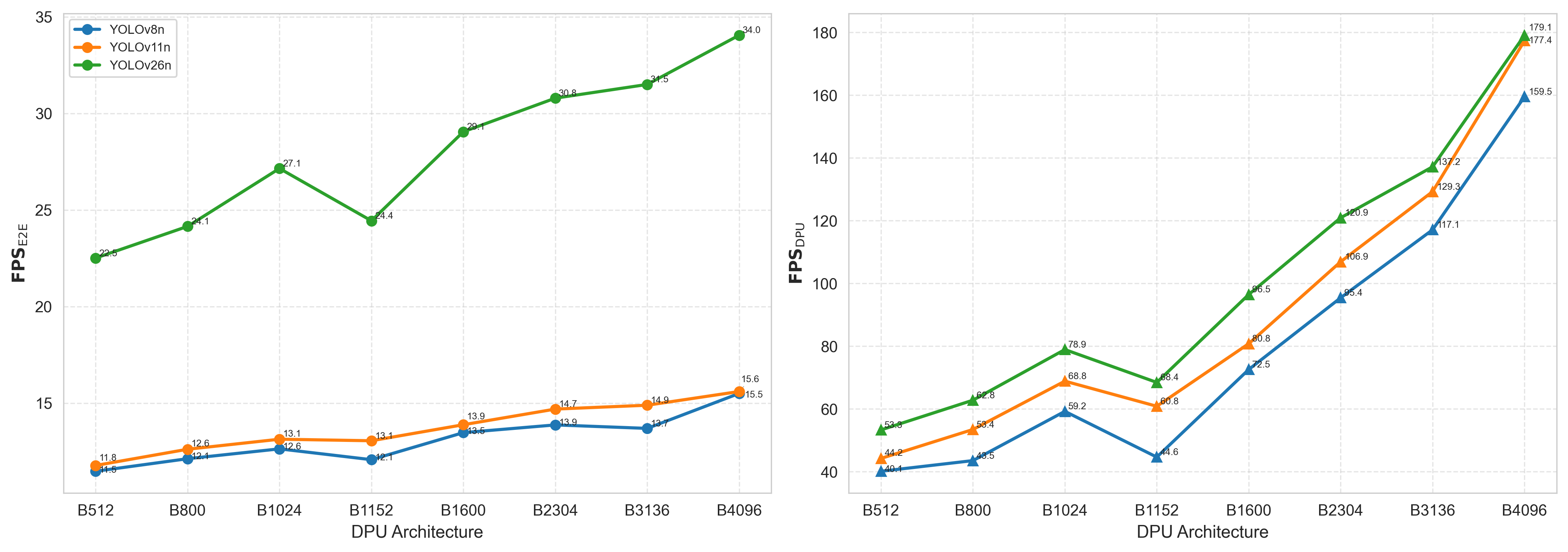}
   \caption{Comparison of end-to-end and DPU FPS for standard object detection models across DPUCZDX8G IP architectures.}
    \label{fig:fps_analysis_standard}
\end{figure*}

\begin{figure*}
    \centering
    \includegraphics[width=\linewidth]{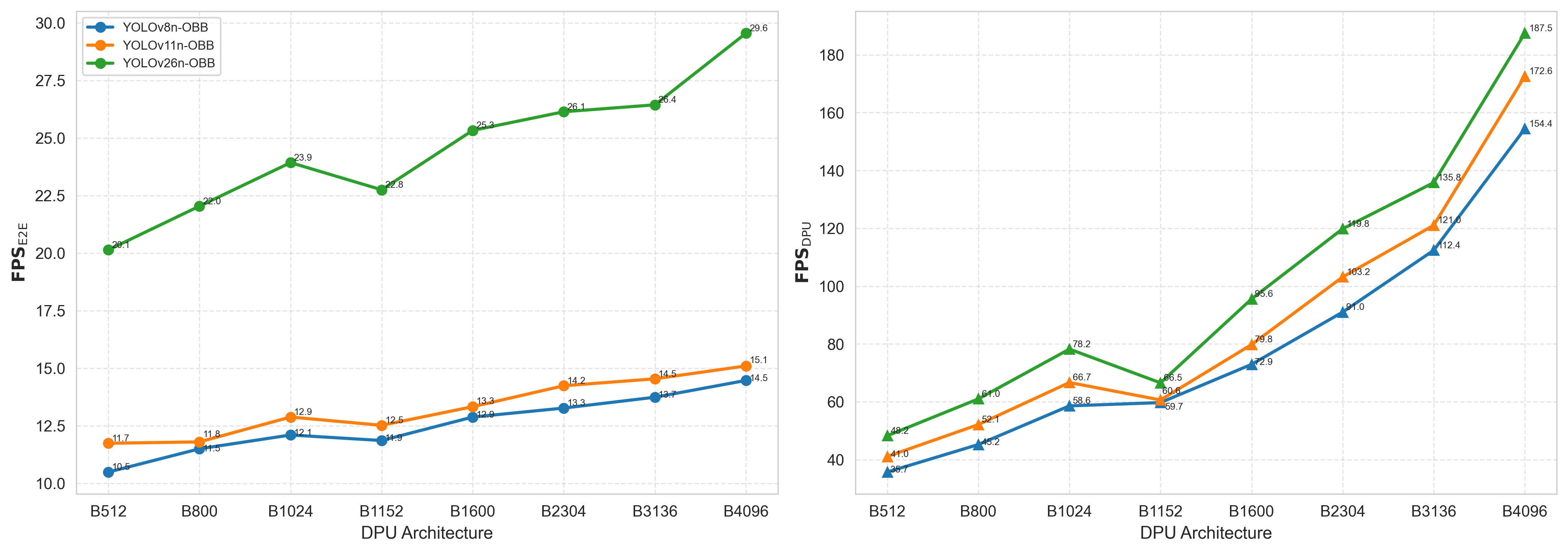}
   \caption{Comparison of End-to-End and DPU FPS for oriented object detection models across DPUCZDX8G IP core architectures.}
    \label{fig:fps_analysis_obb}
\end{figure*}

\begin{figure*}
    \centering
    \includegraphics[width=\linewidth]{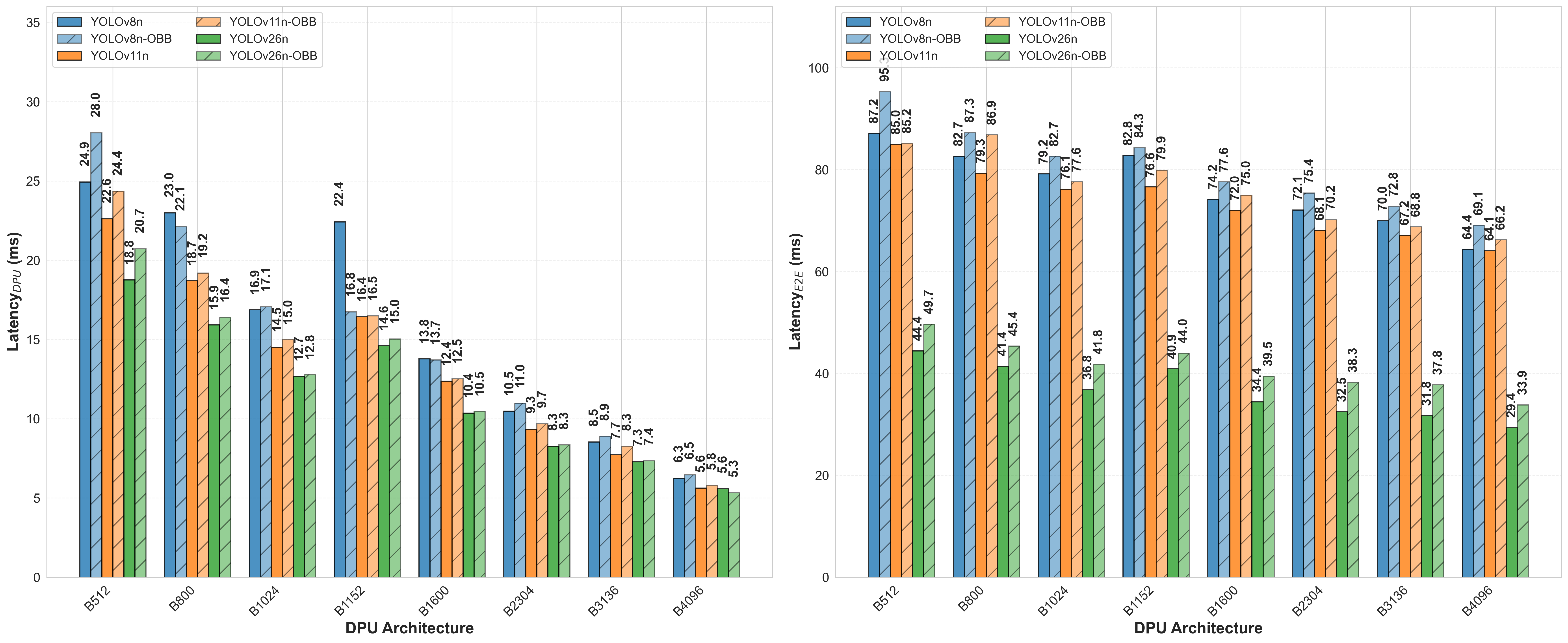}
   \caption{Comparison of end-to-end and DPU latency for standard and oriented object detection models across DPUCZDX8G IP core architectures.}
    \label{fig:latency_comparison}
\end{figure*}

The DPU and end-to-end (E2E) frames per second ($\mathbf{FPS}_{\mathrm{DPU}}$, $\mathbf{FPS}_{\mathrm{E2E}}$) and latency ($\mathbf{Latency}_{\mathrm{DPU}}$, $\mathbf{Latency}_{\mathrm{E2E}}$) of each deployed model are measured on a single DPU core using PETS2009-S2L1 \cite{Ferryman2009PETS2009}, a video with a resolution of $768$×$576$ and a length of $113.57$ seconds ($795$ frames at $7$ FPS), across all available DPUCZDX8G IP architectures. Figures \ref{fig:fps_analysis_standard}, \ref{fig:fps_analysis_obb}, and \ref{fig:latency_comparison} illustrate performance scaling from the B512 to B4096 architectures for standard and oriented object detection models.

As the DPUCZDX8G IP architecture scales up, $\mathbf{FPS}_{\mathrm{DPU}}$ rises and $\mathbf{Latency}_{\mathrm{DPU}}$ decreases consistently for all models, showing that the larger DPUCZDX8G architecture provides significantly higher FPS and lower latency. However, a considerable gap is found between DPU and end-to-end performance in terms of FPS and latency, which indicates a bottleneck in the post-processing step that limits the overall throughput and latency. This entails that upgrading only the architecture of the DPUCZDX8G IP core is insufficient to achieve proportional end-to-end performance. Due to the removal of distribution focal loss (DFL) and architectural optimizations, YOLOv26n and YOLOv26n-OBB achieve the highest end-to-end FPS and lowest latency across all DPUCZDX8G IP core architectures, with $\mathbf{FPS}_{\mathrm{E2E}}$ of $34.05$ and $29.55$, and the lowest $\mathbf{Latency}_{\mathrm{E2E}}$ of $29.37$ ms and $33.85$ ms for standard and oriented object detection when using B4096 DPU, respectively.

\subsection{Resource Utilization}
The power consumption and resource utilization reported by Vivado are shown in Table~\ref{tab:dpu_resource_power}. Both power consumption and resource utilization generally increase with the DPUCZDX8G IP core architecture, from B512 to B4096. The power consumption ranges from $4.792$ W at B512 to $8.760$ W at B4096, representing an increase of approximately $83$\%, while remaining within a reasonable power budget for embedded FPGA deployment. Overall, the B4096 configuration provides the highest parallelism and FPS at the cost of greater resource consumption.

\begin{table*}
\centering
\caption{FPGA resources utilization and On-Chip Power across DPUCZDX8G IP Architectures}
\label{tab:dpu_resource_power}
\setlength{\tabcolsep}{5pt}
\renewcommand{\arraystretch}{1.12}
\begin{adjustbox}{max width=\textwidth}
\begin{tabular}{l c c c c c c}
\toprule
\textbf{DPU} & \textbf{Power (W)} & \textbf{LUTs} & \textbf{BRAM} & \textbf{URAM} & \textbf{DSP} & \textbf{Regs} \\
\midrule
B4096 & $8.760$ & $55.8$k ($24.16$\%) & $142$ ($45.67$\%) & $46$ ($47.92$\%) & $710$ ($41.09$\%) & $111k$ ($21.94$\%) \\
B3136 & $7.864$ & $50$k ($21.74$\%)   & $123$ ($39.58$\%) & $42$ ($43.75$\%) & $566$ ($32.75$\%) & $82k$ ($17.97$\%) \\
B2304 & $7.165$ & $45.1$k ($19.66$\%) & $104$ ($33.49$\%) & $38$ ($39.58$\%) & $438$ ($25.35$\%) & $71k$ ($15.71$\%) \\
B1600 & $6.382$ & $41.1$k ($17.93$\%) & $85$ ($27.40$\%)  & $34$ ($35.42$\%) & $326$ ($18.87$\%) & $61k$ ($13.38$\%) \\
B1152 & $5.571$ & $33.6$k ($14.73$\%) & $38$ ($12.34$\%)  & $38$ ($39.58$\%) & $222$ ($12.85$\%) & $49k$ ($10.78$\%) \\
B1024 & $5.654$ & $36.5$k ($16$\%)    & $76$ ($24.36$\%)  & $15$ ($15.63$\%) & $230$ ($13.31$\%) & $49k$ ($11$\%) \\
B800  & $5.206$ & $31.2$k ($13.71$\%) & $31$ ($10.10$\%)  & $34$ ($35.42$\%) & $166$ ($9.61$\%)  & $42k$ ($9.4$\%) \\
B512  & $4.792$ & $28.2$k ($12.4$\%)  & $28$ ($8.97$\%)   & $15$ ($15.53$\%) & $118$ ($6.83$\%)  & $35k$ ($7.95$\%) \\
\bottomrule
\end{tabular}
\end{adjustbox}
\end{table*}

\section{Conclusion and Future Work}
\label{sec:Con}
This paper presents a systematic benchmarking study of attention-based YOLO models — specifically YOLOv11 and YOLOv26, along with their oriented object detection variants — deployed on an FPGA SoC using the Vitis AI toolchain. Architectural modifications were made, such as replacing the unsupported activation function, substituting split operations with equivalent $1 \times 1$ convolutions, and modifying the spatial attention mechanism. To provide a quantitative baseline for FPGA-based object detection, the models were evaluated across six benchmark datasets — covering both standard and oriented object detection tasks.

The results show that YOLOv26n and YOLOv26n-OBB consistently achieve the highest end-to-end frame rates of $34.05$ and $29.55$, respectively, at the cost of a reduction of approximately $5$\% (absolute) in accuracy compared with CPU execution, representing a common trade-off in edge FPGA deployment.

Power consumption and hardware resource utilization scale up with DPUCZDX8G IP core, ranging from $4.792$W at B512 to $8.760$W at B4096, yet achieve approximately $3{\times}$ lower power consumption compared with the state-of-the-art~\cite{montgomeriecorcoran2023sataystreamingarchitecturetoolflow}.
Notably, post-processing became a critical performance bottleneck due to single-core DPU execution.
This limits end-to-end performance regardless of the DPUCZDX8G IP core architecture used. 
In the future, we will extend this work to adapt the detection system without using NMS, which may result in end-to-end FPS improvement.

\section*{Acknowledgments}
\scriptsize
We gratefully acknowledge funding by the projects enableATO (German Federal Ministry of Transport, German Center for Future Mobility DZM, grant: 19DZ23002D), FH-Personal (German Federal Ministry of Research, Technology and Space (BMFTR), grant: 03FHP106) and KI-Akademie OWL (BMFTR, supported by VDI/VDE Innovation+Technik GmbH, grant: 16IS24057C).

\bibliographystyle{IEEEtran}

\end{document}